\documentclass{desyproc}
\newcommand\ignore[1]{}

\newcommand\be{\begin{equation}}
\newcommand\ee{\end{equation}}
\newcommand\bea{\begin{eqnarray}}
\newcommand\eea{\end{eqnarray}}

\newcommand{\bdm}{\begin{displaymath}}
\newcommand{\edm}{\end{displaymath}}
\newcommand\nn{ \nonumber\\}

\renewcommand{\>}{\rangle}

\begin{document}
\title{Holographic Double Diffractive Production of Higgs\\
and the AdS Graviton/Pomeron}


\author{{\slshape Richard Brower$^1$, Marko Djuri\'c$^2$, Chung-I Tan$^3$\footnote{Speaker}}\\[1ex]
$^1$Physics Department, Boston University, Boston, MA  02215, USA.\\
$^2$Centro de F\'isica do Porto, Universidade do Porto, 4169-007 Porto, Portugal.\\
$^3$Physics Department, Brown University, Providence, RI 02912, USA.}

\contribID{smith\_joe}


\maketitle

\begin{abstract}
 The holographic approach to double diffractive Higgs production is presented for the AdS graviton/Pomeron of
  Brower, Polchinski, Strassler and Tan~\cite{Brower:2006ea}. The goal is to provide a simple framework from the 
dual strong coupling point of view, which nonetheless is capable of providing phenomenologically compelling estimates of the cross sections.  This article is the first step in defining the building block  in anticipation of experimental observations at the LHC.  As in the traditional weak coupling approach in order to constrain the phenomenological parameters, we anticipate  the holographic parameterizations must  subsequently be tested and calibrated through factorization for a self-consistent description of  other diffractive process such as total cross sections, deep inelastic scattering and heavy quark production in the central region.
\end{abstract}

\section{Introduction}

\label{sec:intro}

A promising production mechanism for Higgs meson at the LHC involves the forward proton-proton scattering $p p \rightarrow p H p$. The protons scatter through very
small angles with a large rapidity gaps separating the Higgs in the
central region. The Higgs subsequently decays into large transverse momentum
fragments. Although this represents a small fraction of the total
cross section, the exclusive channel should provide an exceptional
signal to background discrimination by constraining the Higgs mass to both the energy of
decay fragments and the energy lost to the forward protons~\cite{Kharzeev:2000jwa}. Relaxing the kinematics to allow
for inclusive double diffraction may also be useful, where one or both
of the nucleon are diffractively excited. While double diffraction is
unlikely to be a discovery channel, it may play a useful role in
determine properties of the Higgs after discovery.

Current phenomenological estimates of the diffractive Higgs production
cross section have generally followed two approaches:  perturbative (weak coupling) vs  confining (strong coupling), or equivalently, in the Regge literature,  often referred to as the ``hard Pomeron''  vs``soft Pomeron''  methods. The Regge approach to high energy scattering, although well motivated phenomenologically, has suffered in the past by the lack of a precise theoretical underpinning. The advent of AdS/CFT has dramatically changed the situation. In a holographic approach,  the Pomeron is a well-defined concept and it can be identified as the ``AdS graviton" in the strong coupling~\cite{Brower:2006ea}, or, simply the BPST Pomeron. In this talk, we briefly review the general properties of the BPST Pomeron and then show how it can be used to describe double-diffractive production of Higgs.

\section{Holographic Model for Diffractive Higgs Production}\label{sec:Ingredient}

\begin{figure}[h]
\qquad
\includegraphics[height=0.15 \textwidth,width=0.4\textwidth]{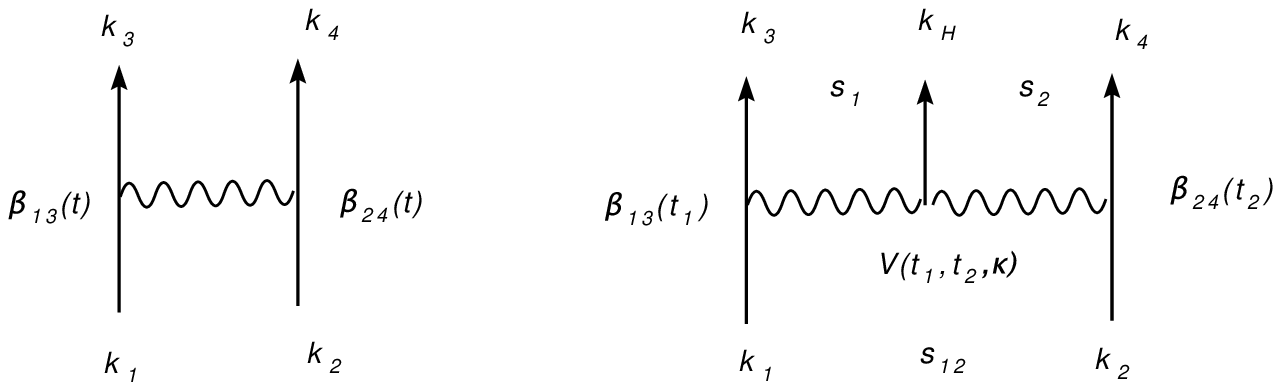}
\qquad
\qquad
\qquad
\includegraphics[angle = 90, height = 0.15\textwidth, width = 0.25\textwidth]{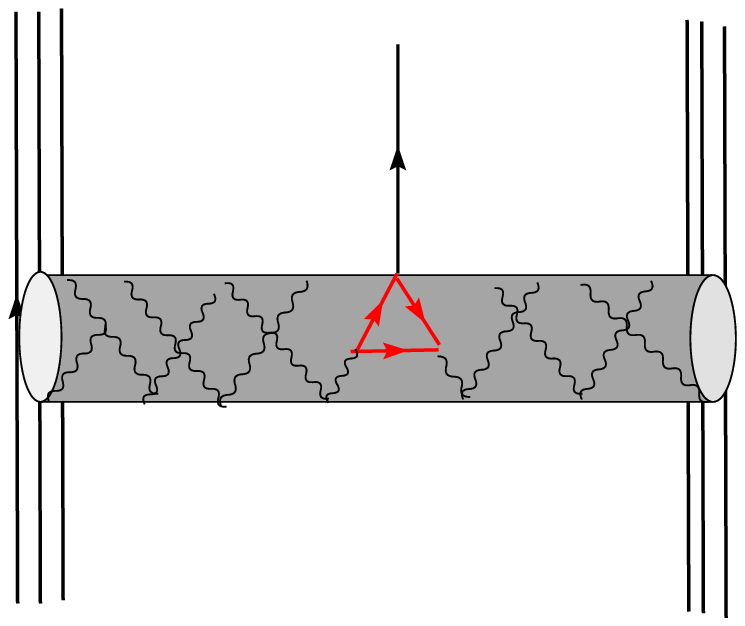}

 \caption{(a) Kinematics for single-Regge limit for  2-to-2 amplitudes, (b) Double-Regge kinematics for  2-to-3 amplitudes.  (c) Cylinder Diagram for large $N_c$ Higgs Production.}
 \label{fig:cylindarHiggs}
\end{figure}

The formulation of AdS/CFT  for high energy diffractive collision
 has already a rather extensive literature to draw on~\cite{Brower:2007qh,Brower:2007xg,Cornalba:2006xm}.
``Factorization in AdS" has emerged as a {\it universal} feature, applicable to scattering involving both particles and currents.  For instance, for elastic scattering, the amplitude can be represented schematically in a factorizable form, 
\be
A(s,t)  = \Phi_{13}*\widetilde {\cal K}_P * \Phi_{24} \; .
\label{eq:adsPomeronScheme}
\ee
where $ \Phi_{13}$ and $ \Phi_{24}$ represented two elastic vertex couplings, and $\widetilde {\cal K}_P$ is an universal Pomeron kernel~\footnote{Unlike the case of a graviton exchange in $AdS$, this Pomeron kernel contains both real and imaginary parts.}, with a characteristic power behavior at large $s>>|t|$, 
\be
\widetilde {\cal K}_P\sim s^{j_0}\, ,
\label{eq:Regge}
\ee 
schematically represented by Fig. \ref{fig:cylindarHiggs}a. This ``Pomeron intercept", $j_0$, lies in the range  $1<j_0<2$ and is a function of the 't Hooft coupling, $g^2N_c$. The  convolution  in (\ref{eq:adsPomeronScheme}), denoted by the $*$-operation, involves an integration over the AdS location in the bulk. (For more details, see the talk by M. Djuri\'c at this Workshop.)   This formalism 
has also been applied  to give a reasonable account of
the small-$x$ contribution to deep inelastic scattering~\cite{Brower:2010wf}.  In moving from elastic to DIS, one simply replaces $\Phi_{13}$ in   (\ref{eq:adsPomeronScheme}) by appropriate product of propagators  for external currents~\cite{Polchinski:2002jw,Brower:2010wf}.

A holographic treatment of Higgs production  amounts to a generalization of our previous $AdS$ treatment for 2-to-2 amplitudes to one for  2-to-3 amplitudes, e.g., from Fig. \ref{fig:cylindarHiggs}a to Fig. \ref{fig:cylindarHiggs}b. A more refined analysis for Higgs production requires a careful treatment for that depicted in Fig. \ref{fig:cylindarHiggs}c.  A particularly useful paper for the diffractive Higgs analysis
is the prior work by Herzog, Paik, Strassler and
Thompson~\cite{Herzog:2008mu} on holographic double diffractive
scattering. In this analysis, one generalizes (\ref{eq:adsPomeronScheme}) to 2-to-3 amplitude where
\be
A(s,s_1, s_2, t_1, t_2)  = \Phi_{13}*  \widetilde{\cal K}_P*V_H*  \widetilde {\cal K}_P* \Phi_{24} \; ,
\label{eq:adsDoublePomeronScheme}
\ee
schematically represented by Fig. \ref{fig:cylindarHiggs}b. 
However, a new aspect, not addressed in \cite{Herzog:2008mu},  is the issue of scale invariance breaking.   A proper accounting for a non-vanishing gluon condensate $\langle F^2\rangle$ turns out to be a crucial ingredient in understanding the strength of diffractive Higgs production.  

\ignore{
\begin{figure}[bthp]
\begin{center}
\includegraphics[width = 0.5\textwidth]{ExtremeElastic.eps}
\end{center}
\caption{ (a) 2 gluon exchange Low-Nussinov Pomeron at $g^2 N_c
  \rightarrow 0$. (b) The extreme strong coupling  Witten diagram of AdS ``graviton/pomeron'' at $1/g^2 N_c \rightarrow 0$.}
\label{fig:extreme}
\end{figure}
}

Let us first list the assumptions and the corresponding  building blocks required
to develop a model for holographic description of diffractive Higgs production. The basic theoretical steps necessary in order to arrive at (\ref{eq:adsPomeronScheme}) and (\ref{eq:adsDoublePomeronScheme}) are: 

(a)  {\it Diffractive Scattering and QCD at Large $N_c$ Limit}: in this  limit, there is a more precise
definition of the ``bare Pomeron''.  In leading order of the $1/N_c$
expansion at fixed 't Hooft coupling $\lambda = g^2N_c$, diffraction is
given peturbatively by the exchange of a network of gluons with the
topology of a cylinder, corresponding in a confining theory to the
t-channel exchange of a closed string for glueball states.  Such a state can be  identified with the Pomeron.

(b)  {\it From Weak to Strong Coupling}: Prior to AdS/CFT,  property of the Pomeron has been explored  mostly from a perturbative approach. The advent of the AdS/CFT correspondence has provided a firmer foundation from which a non-perturbative treatment can now be carried out. 
 For instance, for elastic scattering,  the 2-to-2 amplitude can be represented by the exchange of  a single graviton, schematically given in a factorizable form, 
\be
A(s,t)  = \Phi_{13}*\widetilde {\cal K}_G * \Phi_{24} \; .
\label{eq:adsGravitonScheme}
\ee
where $ \Phi_{13}$ and $ \Phi_{24}$ represented two elastic vertex couplings to the graviton and $\widetilde {\cal K}_G$ is dominated by the ``$(++,--)$" component of the graviton propagator~\cite{Brower:2007qh}.  Since this corresponds to a spin-2 exchange, the dominant graviton kernel $\widetilde {\cal K}_G$ grows with a integral power, i.e., at fixed $t$, as  $s^2$.  Similarly, double diffractive Higgs production will be dominated by a double-graviton exchange diagram, leading to a similar factorizable expression for the production amplitude
\be
A(s,s_1, s_2, t_1, t_2)  = \Phi_{13}*  \widetilde{\cal K}_G *V_H*  \widetilde {\cal K}_G* \Phi_{24} \; .
\label{eq:adsDoubleGravitonScheme}
\ee
In comparing with (\ref{eq:adsGravitonScheme}),  a new Higgs production vertex $V_H$ is required. The central issue in a holographic description for diffractive Higgs production is the specification of this new vertex $V_H$.

(d)  {\it Confinement}: However, above discussion is purely formal since a CFT has no scale and  one needs to be more precise in defining the Regge limit.  First, in order to provide a particle interpretation, the basic framework is a holographic approximation to the dual
QCD with confinement deformation.
With confinement deformation, the $AdS$ is effectively cutoff. Because of the ``cavity effect", both dilaton and the transverse-traceless metric become massive, leading to an infinite set of massive scalar and tensor glueballs respectively.  In particular, each glueball state can be described by a normalizable wave function $\Phi(z)$ in $AdS$. The weight factor   $\Phi_{ij}$ in the respective factorized representation for the elastic and Higgs amplitudes, (\ref{eq:adsGravitonScheme}) and (\ref{eq:adsDoubleGravitonScheme}), is given by $\Phi_{ij}(z) = e^{-2A(z)} \Phi_i(z)\Phi_j(z)$. In contrast, for amplitudes involving external currents, e.g., for DIS~\cite{Polchinski:2002jw,Brower:2010wf}, non-normalizable wave-functions will be used. 

\begin{figure}[bthp]
\begin{center}
\includegraphics[width = 0.35\textwidth]{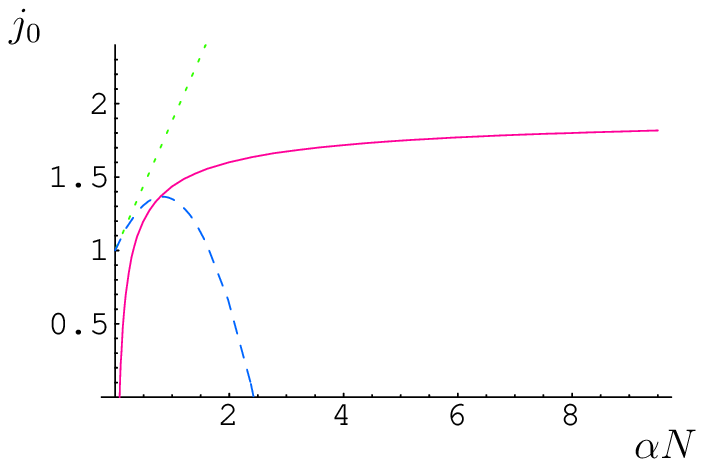}
\end{center}
\caption{In ${\cal N} = 4$ Yang-Mills theory,
the weak- and strong-coupling calculations of the
position $j_0$ of the leading singularity for $t\leq 0$,
as a function of $\alpha N = g^2 N_c/4 \pi$.
Shown are the leading-order BFKL calculation (dotted), the
next-to-leading-order calculation (dashed), and the strong-coupling
calculation of this paper (solid).  Note the latter two
can be reasonably interpolated. }
\label{fig:N4j0}
\end{figure}

(e) {\it Correction to Strong Coupling in $1/\sqrt{\lambda}$}:  It has been shown in \cite{Brower:2006ea}, for ${\cal N} = 4$ SUSY YM, 
the leading strong coupling Pomeron~\cite{Brower:2006ea,Brower:2007qh,Brower:2007xg} is at 
\be
j_0 = 2 - 2 /\sqrt{g^2N_c}    \; .
\label{eq:BPST-intercept}
\ee
which is ``lowered" from  $J =2$ as one decreases $\lambda$.  In a realistic holographic approach to high energy scattering, one must work at $\lambda$ large but finite in order to account for  the Pomeron intercept  of the order $j_0\simeq 1.3$. After taking into account $O(1/\sqrt \lambda)$ correction to the Graviton kernel, one arrives at (\ref{eq:adsPomeronScheme}) and (\ref{eq:adsDoublePomeronScheme}) for elastic and diffractive Higgs production respectively. Here the Pomeron kernel, $ \widetilde{\cal K}_P$,  has hard components due to near conformality in the UV and soft
Regge behavior in the IR.   It is  interesting to compare the weak and strong coupling (conformal) Pomeron by plotting the intercept of the
leading sigularity in the $J-$plane.
 This  is to be compared with the weak coupling
BFKL intercept  to second order, as shown in Fig.~\ref{fig:N4j0}. The phenomenological estimate for 
QCD  gives an intercept of about $j_0 \simeq 1.3$, suggesting
that { the physics of diffractive scattering is roughly in the cross-over region between
strong and weak coupling. }

(f) {\it Higgs Production From Weak   to Strong Coupling}: In a
perturbative approach, often dubbed as ``hard Pomeron", Higgs production can be viewed as gluon
fusion in the central rapidity region.  A Higgs can be produced at central rapidity by the double Regge Higgs vertex through
a heavy quark loop which in lowest order is a simple gluon fusion process, dominant for large parton x for the colliding gluons.
A more elaborate picture emerges as one tries to go  to the region of the softer (wee gluons) building up
double Regge regime~\footnote{In addition to  the Pomeron exchange contribution in these models must
subsequently be reduced by large Sudakov correction at the Higgs
vertex and by so called survival probability estimates for soft gluon
emission, again reflecting the view that double diffraction Higgs
production is intrinsically non-perturbative.}
In the large $N_c$ there are no quark loop in the bulk of AdS space and since
the Higgs in the Standard Model only couples to quark via the
Yukawa interactions there appears to be a problem with strong coupling Higgs production
in leading $1/Nc$.  Fortunately the solution to this is to follow the standard procedure in Higgs phenomenology, which is to
integrate out the quark field replacing the Higgs coupling to the gauge operator $Tr[F^2]$. 

Consider the Higgs coupling to quarks via a Yukawa coupling,  and, for simplicity we will assume is dominated by the top quark. We will be more explicit in the next Section, and simply note here that, 
 after taking advantage of the scale separations between the QCD scale, i.e., the Higgs mass and the top quark mass,   $\Lambda_{qcd} \ll m_H \ll 2 m_t$, 
heavy quark decoupling allows one to replace the Yukawa coupling  by an effective interaction,
\be
{\cal L} = \frac{\alpha_s g}{24 \pi M_W} F^a_{\mu\nu}F^{a \mu \nu} \phi_H
\label{eq:effectiveHiggsCoupling}
\ee
by evaluating the two gluon Higgs triangle graph in leading order $O(M_H/m_t)$. 
Now the AdS/CFT dictionary simply requires that this be the source in the UV of the AdS dilaton field.  It follows, effectively, for Higgs production, we are required to work with a five-point amplitudes, one of the external leg involves a scalar dilaton current coupling to $Tr[F^2]$.  For diffractive Higgs production, in the supergravity limit, the Higgs vertex $V_H$ is given by a two-graviton-dilaton coupling, Fig. \ref{fig:cylindarHiggs}c.

(g) {\it Conformal Symmetry Breaking}: We now must pause to realize that in any conformal theory the is no dimensional 
parameter to allow for such a  dimensionful two-graviton-dilaton coupling, $M^2 \phi h_{\mu\nu} h^{\mu \nu}$,  emerging  in an  expansion of the AdS gravity action  if scale invariance is maintained.  However since  QCD is not a conformal theory 
this is just one of many reasons to introduce conformal symmetry breaking.  Many attempts have
been made to suplement this phenomenological Lagrangian with other fields such as
the  gauge fields for the light quark Goldstone modes to provide
a better holographic dual for QCD.  In principle enen at
leading order of large $N_c$ we should eventually
require  an infinite number of (higher spin) field in the bulk representation
to correspond the yet undiscovered 2-d sigma model for the world-sheet string theory
for  QCD.   Fortunately for the phenomenological
level at high energy, these details are non-essential.  To model an effective QCD background we will for the most
part introduce two modifications of the pure AdS background: (1) an IR hardwall cut-off 
beyond $z = 1/\Lambda_{qcd}$ to give confinement and  linear static quark potential at large distances and (2) 
a slow deformation in the UV ($z \rightarrow 0$) to model the logarithmic running 
for asymptotic freedom.  Both break conformal invariance, which
as we will argue is required to couple the two gravitons to the dilaton and produce a
Higgs in the central rapidity region.

After taking into account of finite $\lambda$ correction, the leading order Higgs production diagram at large $N_c$ can be schematically represented in Fig.~\ref{fig:cylindarHiggs}c, with each of the left- and right-cylinder representing a BPST Pomeron. It should be pointed out, just as in the case  of elastic scattering, it is necessary to consider higher order corrections, e.g., eikonal corrections. We will not do it here, but will address this issue in the conclusion section.  In what follows, we shall focus on the Pomeron-Pomeron fusion vertex in the strong coupling limit.

Finally it should be noted that
one critical missing ingredient of these ad hoc conformal breaking deformation of
the AdS geometry  in the UV and IR  is the
fact the spontaneous  breaking of pure Yang Mills ( and presumable  QCD
at large $N_c$), via ``dimensional transmutation'' 
eliminates the coupling, $\lambda$, as a free parameter. It is fixed 
via the beta function in terms of a single integration constant (sometime called  $\Lambda_{qcd}$ ) which provides the only mass scale.  Thus the logarithmic scale violation in the UV are tied to the same parameter giving confinement in the IR. All holographic modes of QCD to date introduce two mass scales and thus neglect this constraint. The solution to this problem also presumably awaits the
determination of the unique string theory for large $N_c$ QCD.

\section{Pomeron-Pomeron fusion Vertex}
\label{sec:PPfusion}

We are now in a position to focus on the issue of double diffractive
Higgs production from the perspective of String/Gauge duality, i.e.,
the Higgs vertex, $V_H$. It is important to stress that our general discussion  in moving from
single-Pomeron exchange processes, (\ref{eq:adsPomeronScheme}), to double-Pomeron exchange, (\ref{eq:adsDoublePomeronScheme}), applies equally well for both
diffractive glueball production and for Higgs production. The
difference lies in how to treat the new central vertex.  For the
production of a glueball, the vertex will be proportional to a
normalizable $AdS$ wave-function.  There will also be an overall factor
controlling the strength of coupling to the external states, e.g., the
Pomeron-Pomeron-glueball couplings.  For Higgs production, on the
other hand, the central vertex, $V_H$, involves a non-normalizable
bulk-to-boundary propagator, appropriate for a scalar external
current.This in turns leads to coupling to a Higgs scalar.  The
difference between these two cases parallels the situation for
four-point amplitudes in moving from proton-proton (p-p) elastic
scattering to electron-proton deep-inelastic scattering (e-p DIS). In
moving from p-p to DIS, one simply replaces one of the two pairs of
normalizable proton wave-functions  with a
pair of non-normalizable counterparts appropriate for conserved
external vector currents.

A Higgs scalar in the standard model couples exclusively to the quarks
via Yukawa coupling, which for simplicity we will assume is dominated
by the top quark, with   
\be
{\cal L} = - \frac{g}{2 M_W} m_t \; \bar t(x) t(x) \phi_H(x).
\label{eq:HiggsCoupling2ttbar}
\ee
 Taking advantage of the scale separations between the QCD scale, the Higgs mass and the top quark mass,
$
\Lambda_{qcd} \ll m_H \ll 2 m_t
$,
heavy quark decoupling allows one to replace the Yukawa coupling by direct coupling of Higgs to gluons, which is treated  as an external source in the AdS dictionary.
Consequently $V_H$, in a coordinate representation,  is replaced by the vertex  for two AdS Pomerons fusing
at $(x'_{1\perp},z'_1)$ and $(x'_{2\perp},z'_2)$ and propagating this
disturbance to the $\bar t(x) t(x)$ scalar current at the boundary of
AdS.  The double diffractive Higgs vertex $V_H$ can then be obtained in a two-step process.

First, since the
Yukawa Higgs quark coupling is proportional to the quark mass, it is
dominated by the top quark. Assuming  $m_H \ll m_t$,  this can be
replaced by an effective interaction, (\ref{eq:effectiveHiggsCoupling}),
by evaluating the two gluon Higgs triangle graph in leading order $O(M_H/m_t)$.
Second, using the AdS/CFT dictionary, the external
  source for $F^a_{\mu \nu}F^a_{\mu \nu}(x)$ is placed at the
  AdS boundary ($z_0 \rightarrow 0$) connecting to the Pomeron fusion vertex
in the interior of $AdS_3$ at ${\bf b}_H=(x'_H,z'_H)$, by a scalar bulk-to-boundary propagator,
$K(x'_H - x_H,z'_H,z_0)$.

We are finally in the position to put all the pieces together. Although we eventually want to go to a coordinate representation in order to perform eikonal unitarization, certain simplification can be achieved more easily in working with the  momentum representation. The Higgs production amplitude, schematically given by (\ref{eq:adsDoublePomeronScheme}), can then be written explicitly as
 \bea
A(s,s_1,s_2, t_1,t_2)&\simeq &  \int dz_1 dz  dz_2\; \sqrt{-g_1}\sqrt{-g} \sqrt{-g_2}\;\Phi_{13} (z_1)     \nn
&\times&   \widetilde {\cal K}_P(s_1,t_1,z_1,z)  \; V_H(q^2 , z)\; \widetilde {\cal K}_P(s_2,t_2,z,z_2)\; \Phi_{24}(z_2)  \;.
\label{eq:adsDoublePomeronHiggs}
\eea
where  $q^2= -m_H^2$.   For this production vertex,  we will keep  it simple by  expressing it as 
\be
 V_H( q^2,z)= V_{PP\phi} K(q^2,z) L_H\; .
 \label{eq:HiggsVertex}
\ee
where  $K(q^2,z)$ is the conventionally normalized bulk to boundary propagator, $V_{PP\phi}$ serves as  an overall coupling from two-Pomeron to $F^2$, and $L$ is  the  conversion factor from $F^2$ to Higgs, i.e.,
$
L_H = L(-m_H^2)\simeq  \frac{\alpha_s g}{24 \pi M_W}.
$
By   treating the central vertex $  V_{PP\phi}$ as a constant, which follows from the super-gravity limit, we have ignored possible additional dependence on  $\kappa$, as well as that on $t_1$ and $t_2$. This approximation gives an explicit
factorizable form for Higgs production.

\section{Strategy for Phenomenological Estimates}
\label{sec:strategy}

While we  intend to lay in  this article the formal framework for the holographic diffractive Higgs production approach, it is useful to outline the phenomenological approach we plan to pursue to confront experimental data. There should be a strong warning however that details will necessarily change as we discover which parameterization are critical to a global analysis of data. Our current version for the holographic Higgs amplitude involves 3  parameters: (1) the IR cut-off determined by the glueball mass, (2) the leading singularity in the $J$-plane determined~\footnote{In a true dual to QCD, there is no independent parameter for   the strong coupling, because of ``dimensional transmutation'', which fixes all dimensionful quantities   relative to the a single mass scale $\Lambda_{qcd}$, through the running   coupling constant. For instance, the glueball mass in units of $\Lambda_{qcd}$ is fixed and   computed in lattice computations. } by the 't Hooft parameter $g^2N_c$ and (3) the strength of the central vertex parameterized by the string coupling or Planck mass. A strategy must be provided in fixing  these parameters.

As a first step in this direction, we ask how the central vertex, $V_H$, or equivalently,  $ V_{PP\phi}$,  via (\ref{eq:HiggsVertex}),  can be normalized, following  the approach of Kharzeev and Levin~\cite{Kharzeev:2000jwa} based on the analysis of trace anomaly. 
 We also show how
one can in principle use the elastic scattering to normalize the bare BPST Pomeron coupling to external protons and the 't Hooft 
coupling $g^2N_c$. As in the case of elastic scattering, it is pedagogically reasonable to begin by first treating the simplest case  of double-Pomeron exchange for Higgs production, i.e., without absorptive correction. We discuss how phenomenolgically reasonable simplifications can be made. This is followed by treating eikonal
corrections in the next section, which provides a means of  estimating the  all-important survival probability.

\subsection{Continuation to Tensor Glueball Pole and On-Shell Higgs Coupling:}

Confinement deformation in AdS will lead to glueball states, e.g., the lowest tensor glueball state lying on the leading Pomeron trajectory~\cite{Brower:2000rp}.   There will also be scalar glueballs associated with the dilaton.  With scalar invariance broken, this  will also lead to non-vanishing couplings between a pair of tensor glueballs and scalar glueballs. 
In terms of the language of Witten diagram,  corresponds  to a non-vanishing graviton-graviton-dilaton coupling in the bulk, which  in turn leads to  $V_H\neq 0$.

Consider first the elastic amplitude. With confinement,
each Pomeron kernel  will contain a tensor glueball pole when $t$ goes on-shell. Indeed, the propagator for our Pomeron kernel can be expressed as a discrete sum over pole contributions.   That is, when $t\simeq  m_0^2$, where $m_0$ is the mass of the lightest tensor glueball, which lies on the leading Pomeron trajectory.  In this limit, the elastic amplitude then  takes on the expected  pole-dominated  form,
\be
A(s,t) \simeq g_{13}\; \frac {s^2}{m^2_0-t} \; g_{24}
\label{eq:2to2KL}
\ee
with vertex $g_{ij}$  given by an overlapping integral:
$
g_{ij}(m_0^2)=\alpha'   \int dz \sqrt{-g(z)} e^{-4A(z)}  \;\beta(m_0^2) \Phi_i(z)\Phi_j(z) \; \phi_{G}(z).
$
Here   $\phi_{G}(z)$ is the wave function for the tensor glueball. 
We have also generalized   $\Phi_{ij}$   by writing it as $\Phi_{ij} (t,z)=\beta(t) e^{-2A(z)}\Phi_i(z)\Phi_j(z)$ for phenomenological  reasons.  That is, the external coupling  $g_{ij}$ is given by an overlap-integral over a product of three wave functions,   $\Phi_i(z)$, $ \Phi_j(z)$  and $  \phi_{G}(z)$. With the standard normalization, 
 $A(s,t)$ is dimensionless.

A similar analysis can also be carried out for   the Higgs production amplitude, Eq. (\ref{eq:adsDoublePomeronHiggs}). Note that the Pomeron kernel now appears twice, $\widetilde {\cal K}_P(s_1,t_1,z_1,z) $ and $\widetilde{\cal K}_P(s_2,t_2,z_2,z) $.   When nearing the respective tensor poles at $t_1\simeq  m_0^2$ and $ t_2\simeq  m_0^2$,  the amplitude can be expressed as
\bea
A(s,s_1,s_2, t_1,t_2)&\simeq&g_{13}\; \frac{  \Gamma_{GGH} \;  s^2  }{(t_1-m_0^2)(t_2-m_0^2)  }   \; g_{24}
   \label{eq:2to3KL}
\eea
As for the elastic case, we have performed the $z_1$ and $z_2$ integrations,  and   have also made use of the fact that $s_1s_2\simeq \kappa\; s\simeq m_H^2 s $. 
Here $\Gamma_{GGH}$ is the effective on-shell glueball-glueball-Higgs coupling, which  can also be expressed as
\be
 \Gamma_{GGH} =L_H F ( - m^2_H)
 \ee
where $L_H=    \frac{\alpha_s g}{24 \pi M_W}$ and     $ F $  is a scalar form factor
$
F(q^2) = \<G, ++,q_1| F^a_{\mu\nu} F^a_{\mu\nu}(0) | G,--,q_2\>.
$ 
That is, in the high energy Regge limit, the dominant contribution comes from  
 the maximum helicity glueball state~\cite{Brower:2006ea}, with $\lambda = 2$.  In this limit,  this form factor,  is given by
the overlap of the dilaton bulk to boundary propagator
\bea
F(q^2) &=&(\alpha' m_H^2)^2   V_{PP\phi} \int dz \sqrt{-g(z)} e^{-4A(z)} \phi_G(z) K(q,z) \phi_G(z)
\label{eq:formfactor}
\eea
  What remains to be specified is the overall normalization, $F(0)$.

We next follow  D. Kharzeev and E. M. Levin  \cite{Kharzeev:2000jwa}, who noted  that, from the SYM side,  $F(q^2)$ at $q^2=0$, can be considered as the glueball condensate. Consider matrix elements of  the trace-anomaly  between two states, $|\alpha(p)\>$ and $|\alpha'(p')\>$, with four-momentum transfer $q=p-p'$. In particular, for a single particle state of a tensor glueball $|G(p)\>$, this leads to $\<G(p)|  {\Theta}^\alpha_\alpha|G(p')\> = \frac{\widetilde \beta}{2 g} \<G(p)| F^a_{\mu \nu} F^{a \mu \nu} | G(p')  \>$. 
At $q = 0$, the forward matrix element of the trace of the energy-momentum tensor is given simply by  the mass of the relevant tensor glueball, with  $\<G|{\Theta}^\alpha_\alpha | G  \> = M_G^2$, this directly yields
\be
F(0) = \<G| F^a_{\mu \nu} F^{a \mu \nu} | G  \>  =-  \frac{4\pi  M_G^2}{3 \widetilde \beta }
\label{eq:FF}
\ee
where $\widetilde \beta = - b\alpha_s/(2 \pi)$, $b = 11 - 2n_f/3$, for $N_c=3$.  In what follows, we will use $n_f=3$.   Note that heavy quark contribution is not included in this  limit.  Since the conformal scale breaking is due the running coupling constant in QCD, 
there is apparently a mapping between QCD scale breaking and breaking of
the AdS background  in the IR, which  gives a finite mass to the glueball
and to give a non-zero contribution to the gauge condensate.

\subsection{Extrapolation to  the Near-Forward limit:}

To apply the above result to the physical region, one needs to extrapolate from $t$ near the tensor pole to the physical region where $t\simeq 0$. Let us first treat the elastic amplitude. A key difference, as one moves from $t\simeq m_0^2$ to $t\simeq 0$,  is the fact that the amplitude becomes complex, with the leading $s$-dependence slowing down from $s^2$ to $s^{j_0}$,  $1< j_0<2$.  To carry out this analysis, it is necessary to result to the $J$-plane representation for the Pomeron kernel $\tilde {\cal K}_P(s,t,z,z')$, with the $J$-plane propagator  $\widetilde G_j(t,z,z')$ given by  a sum of $J$-plane poles, e.g., for hardwall model. For our current purpose, it is sufficient to keep the  contribution coming  from the effective leading trajectory, 
\begin{equation}
\widetilde G_j(z,z';t)\simeq  \widetilde \phi_{eff}(z,j)   \frac{  1 }{m^2_{eff}(j) -t}   \widetilde \phi_{eff}(z',j) \;.
\label{eq:leadingpole}
\end{equation}
where we approximate the  BPST-cut contribution by that of an effective leading pole,  with  the Pomeron kernel behaving as $s^{j_{eff}(t)}$, where  $j_{eff}(t)$, the trajectory function, determined by $m^2_{eff}(j(t)) = t$. That is, we assume $j_{eff}(t)$  remains real in the physical region where $t<0$. By performing the inverse Mellin transform, the large $s$-behavior of the BPST kernel 
can easily be obtained, leading to 
\begin{equation}
A(s,t) \simeq g_{13}(t)\;\left( \frac{ \xi(j_{eff}(t))\; ( \alpha' s)^{j_{eff}(t)}}{\alpha'^{2}\widetilde m^2(t)  }\right)  \; g_{24}(t)
\label{eq:2to2P}
\end{equation}
where   $\widetilde m^2$ is the inverse of trajectory slope, $\widetilde m^2(t) \equiv  dm^2_{eff}(j(t))/dj$, and  $\xi(j)$ is the signature factor.
For the elastic amplitude, the coupling
\begin{equation}
g_{ij}(t)= \alpha'  \int dz \sqrt{-g(z)}  \;\Phi_{ij}(t,z) e^{-j_{eff}(t)A(z)} \widetilde \phi_{0}(z,j_{eff}(t))
\end{equation}
is again in the form of an overlapping integral over the product of three wave functions, with $ \widetilde \phi_{eff}=\widetilde \phi_{0}(z,j_{eff}(t))$.  This serves as a continuation away from the on-shell spin-2 exchange by  replacing the spin-2 wave function $\phi_G(z)= \widetilde \phi_0(z,2)$ by a corresponding wave function for a Pomeron, $ \widetilde \phi_{0}(z,j_{eff}(t))$, with spin shifted from 2 to $j_{eff}(t)$.  Although this shift is of the order $O(1/\sqrt \lambda)$, it is important to note that
 $\widetilde \phi_0(z,j(t))\sim z^{\Delta(j(t))-2} $, for $z\rightarrow 0$,  in contrast to  $\widetilde \phi_0(z,2) \sim z^{2}$. Note that  we have continued with the convention where $g_{ij}(t)$ has  the dimension of length.

As one further continues to the physical region where $t\leq 0$, the amplitude will now be dominated by the contribution from the BPST cut, with  the inverse Mellin transform in $J$ turning into an integral over the discontinuity across the cut,  $(-\infty, j_0)$. Since the contribution from a cut is no longer factorizable, it leads to diffusion in the AdS-radius.    Analytic expression is available in the conformal limit, and, with a hardwall,  a similar analysis can also be carried out by  parametrizing the leading Regge singularity, e.g.,  $m(j)\simeq m_0+ (m_0-m_1)\left(\sqrt {\sqrt\lambda(j-j_0)/2}-1\right)$, leading to  our AdS representation for the elastic amplitude in the forward limit, Eq. (\ref{eq:adsPomeronScheme}).

\subsection{First Estimate for Double-Pomeron Contribution to Differential Cross Section}
 \label{sec:guess}
For our present purpose, it is adequate to first  ignore diffusion by adopt a simpler ansatz  for the elastic amplitude by freezing the AdS radius at $z_0\sim 1/\Lambda_{QCD}$. That is, we assume, for protons,  wave-functions are concentrated near $z_0$, we can replace the Pomeron kernel  with $z$ and $z'$ evaluated at $z_0$, i.e.,
$
\tilde {\cal K}_P(t,s,z,z')  \rightarrow 
\tilde {\cal K}_P(t,s,z_0,z_0)
$
with the resulting $z$ and $z'$ integration leading to unity. 

Focusing next on the forward limit $t=0$, we denote the effective intercept by $\bar j_0$ and  inverse slope by $\widetilde m^2$. Together with the forward coupling $g_{ij}(0)$, 
they will be determined phenomenologically.  We note that $\widetilde m^2$ can be chosen to be  of the order of  the tensor glueball mass, $m_0^2$. For consistency, we also assume that $\bar j_0\simeq j_0$. A corresponding  treatment at $t_1\simeq t_2\simeq 0$ for  the Higgs production amplitude, Eq. (\ref{eq:adsDoublePomeronHiggs}), can lead to a similar simplification. 
   It follows,  after a bit of algebra,
   \bea
A(s,s_1,s_2, t_1\simeq 0, t_2\simeq 0)&\simeq &  g_{13}(0)\frac{  \xi(\bar j_0)^2 \Gamma_{PPH}\; ({\alpha' } s)^{\bar j_0}    }{(\alpha' \widetilde m^2)^2} \     \;     g_{24}(0)
\eea
with an effective central vertex, related to $V_{PP\phi}$ by
\bea
\Gamma_{PPH}&\simeq& \frac{\alpha_s g}{24 \pi M_W}  V_{PP\phi} \left({\alpha' m_H^2}\right)^{\bar j_0}C(\bar j_0) 
\eea
where 
\begin{equation}
C(\bar j_0) = \int dz \sqrt{-g} e^{-4A(z)} \widetilde \phi_0(z,\bar j_0) K(-m^2_H,z) \widetilde \phi_0(z,\bar j_0)
\label{eq:C0}
\end{equation}
and we have dropped terms lower order in $O(1/\sqrt\lambda)$. 
We point out  that (\ref{eq:C0})  is finite due to the wave-function normalizability. For hard-wall, it is  logarithmically divergent as $\bar j_0\rightarrow j_0$ which corresponds to  the onset of a Regge cut. In  a proper treatment when the leading singularity is  a cut, this apparent divergence will be absent. In order to avoid complicating the discussion, we proceed with the understanding that $C(\bar j_0)$  is of the order unity.


Let us turn next to the non-forward limit.  We  accept the fact that, in the physical region where $t<0$ and small, the cross sections typically have an exponential form, with a logarithmic slope which is mildly energy-dependent. We therefore approximate all amplitudes in the near forward region where $t<0$ and small,
$
A(s,t) \simeq  e^{B_{eff}(s)\; t/2}\;  A(s,0)
$ 
where $B_{eff}(s)$ is a smoothly slowly increasing function of s, (we expect it to be logarithmic).
We also assume, for $t_1<0$, $t_2<0$ and small, the Higgs production amplitude is also strongly damped so that
\be
A(s,s_1,s_2, t_1, t_2)  \simeq    e^{B'_eff(s_1) \; t_1/2}  e^{B'_eff(s_2)\; t_2 /2   }  \; A(s,s_1,s_2, t_1\simeq 0, t_2\simeq 0)
\label{eq:higgs}
\ee
We also assume $B'_{eff}(s)\simeq B_{eff}(s) + {\rm  b}$.   With these, both the elastic, the total pp cross sections and the Higgs production cross section can now be evaluated.  Various cross sections will of course depend on the unknown slope parameter, $B_{eff}$, which can at best be estimated based  on prior experience with diffractive estimates.

The phase space for diffractive Higgs production can be specified by the rapidity of Higgs $y_H$, and two-dimensional transverse momenta $q_{i,\perp}$,  $i=3,4,5$, with $q_{5,\perp}=q_{H,\perp}$, in a frame where the incoming momenta $k_1$ and $k_2$ are longitudinal.   Alternatively, due to momentum conservation, we can use instead $y_H, t_1, t_2, \cos \phi$ as four independent variables where  $t_1\simeq - q_{3,\perp}^2$,  $t_2\simeq - q_{4,\perp}^2$, and $\cos \phi=\hat q_{3,\perp}\cdot \hat q_{4,\perp}$.   However, the amplitude is effectively independent of $\phi$ since its dependence enters through the $\kappa$ variable where $\kappa\simeq m_H^2+ q_{H,\perp}^2= m_H^2 + (q_{2,\perp}+q_{4,\perp})^2$. As  discussed earlier, for Higgs production, we can replace $\kappa$ by $\kappa_{eff} \simeq m_H^2$.

Following the earlier analysis,  it is now possible to provide a first estimate for the double-diffractive Higgs production. It is possible to  adopt an approach advocated in by Kharzeev and Levin where the dependence on $B_{eff}$ can be re-expressed in terms of other physical observables. Under our approximation, it is easy to show that  the ratio $ \sigma_{el}/\sigma^2_{total}$ can be expressed as
$
 \frac{\sigma_{el}}{\sigma^2_{total}}=\frac{ 1+\rho^2 }{16\pi B_{eff} (s) }
$ 
where
$
\rho\equiv {{\rm Re } \; {\cal K}(0,s,z_0,z_0) }/{  {\rm Im }\;  {\cal K}(0,s,z_0,z_0) }.
$
Equivalently, one can relate $B_{eff}$ directly in terms of the experimentally smooth dimensionless ratio,
$
R_{el}(s) = {\sigma_{el}}/{\sigma_{total}}  = \frac{ (1+\rho^2) \sigma_{total}(s) }{16\pi B_{eff} (s) }.
$
Upon squaring the amplitude, $A(s,s_1,s_2, t_1, t_2)$, (\ref{eq:higgs}), the double-differential cross section for Higgs production can now be obtained. 
 After integrating over $t_1$ and $t_2$ and using the fact that, for $m_H^2$ large  $s \simeq  {s_1 s_2}/{m_H^2}$,
one finds
\bea
\frac{d\sigma}{dy_H } &\simeq &(1/\pi) \times C'   \times |   \Gamma_{GGH}(0)/\widetilde m^2|^2  \times \frac{\sigma( s)}{\sigma(m_H^2)} \times R^2_{el}(m_H \sqrt s )
\eea
In this expression above, both $C'$ and $\widetilde m^2$, like  $ m_0^2$,  are model dependent. It is nevertheless interesting to note that, since $\Gamma_{GGH}(0)\sim m_0^2$, the glueball mass scale also drops out, leaving a model-dependent ratio of order unity.  In deriving the result above, we have replaced $B'_{eff}$ by $B_{eff}$ where the difference is unimportant at high energy.  With $m_H$ in the range of $100 GeV$, $R_{el}$ can be taken to be  in the range $0.1$ to $0.2$. For $C'\simeq 1$, we find  $\frac{d\sigma}{dy_H }\simeq .8\sim 1.2  \;\; {\rm pbarn}$.
This is of the same order as estimated in \cite{Kharzeev:2000jwa}.  However, as also pointed in \cite{Kharzeev:2000jwa}, this should be considered as an over-estimate. The major source of suppression will come from absorptive correction, which can lead to a central production cross section in the  femtobarn  range. We turn to this next.

\section{Discussion}
\label{sec:discusion}

We conclude by  discussing how consideration of higher order contributions via an eikonal treatment leads to corrections for the central Higgs production. Following by now established usage, the resulting production cross section can be expressed in terms of a ``survival probability".

 Although the ``bare Pomeron'' approximation dominates in
the large $N_c$ expansion, it is clear that higher order summations are
necessary in order to restore unitarity.    In flat space Veneziano
has shown that higher closed string loops for graviton scattering
eikonalize. Indeed in Refs. \cite{Brower:2007qh,Brower:2007xg} it was shown that the same sum leads to an eikonal
expansion that exponentiates for each string bit frozen  in impact parameter
during the collision. To be more explicit, the resulting  eikonal sum leads
to an impact representation for the 2-to-2 amplitude
\bea
A(s,x^\perp - x'^\perp)&=& - 2i s \int
 dz \ dz'\ P_{13}(z) P_{24}(z')  \left [
e^{i\chi(s,x^\perp - x'^\perp, z,z')} - 1\right]
\label{eq:adseik1}
\eea
  The eikonal $\chi$, as  a function of $x_\perp- x'_\perp$, $z,
z'$ and $ s$, can be determined by matching the first order term in $\chi$
to the
single-Pomeron contribution.
In impact space representation, and one finds
$
\chi(s,x_\perp- x'_\perp,z,z')= \frac{g^2_0 }{2 s} \; \widetilde {\cal  K}(s,x_\perp- x'_\perp, z,z')
$


This eikonal analysis can be extended directly to Higgs production. To  simplify the discussion, we shall adopt  a slightly formal treatment. Since  Higgs  is not part of  the  QCD dynamics, one can formally treat our eikonal as a functional of a
weakly coupled  external background Higgs field, $\phi_H(q^\pm,x^\perp_H,z_H)$, that is,  in (\ref{eq:adseik1}), we replace $A(s,x_\perp,x'_\perp)$ and    $\chi(s,x^\perp - x'^\perp, z,z')$ by $A(s,x^\perp, x'^\perp; \phi_H)$ and  $\chi(s,x^\perp - x'^\perp, z,z'; \phi_H)$,   with the understanding that they  reduce to $A(s,x_\perp,x'_\perp)$ and $\chi(s,x^\perp - x'^\perp, z,z')$ respectively in the limit $\phi_H \rightarrow 0$.  Since Higgs production is a small effect, by expanding to first order in the Higgs
background field, we find the leading order Higgs production amplitude, to all order in $\chi$, becomes
\bea
A_H(s_1, s_2,x^\perp -
x^\perp_H,x'^\perp -x^\perp_H, z_H) &=& 2 s \int
 dz \ dz'\ P_{13}(z) P_{24}(z')  \nn
 &\times&  \chi_H(s_1, s_2,x^\perp -
x^\perp_H,x'^\perp -x^\perp_H, z,z',z_H) \;    e^{i\chi(s,x^\perp - x'^\perp, z,z')}\nn
\eea
where $  \chi_H $ can be found  by matching in the limit $\phi_H\rightarrow 0$ with the Higgs production amplitude, (\ref{eq:adsDoublePomeronHiggs}),  due to double-Pomeron exchange in an impact representation. The net effect of eikonal sum is to introduce a phase factor 
$
e^{i\chi(s,x_\perp- x'_\perp,z,z')} $
 into the production amplitude. Due to its absorptive part, ${\rm Im}\; \chi >0$, this eikonal factor provides a strong suppression for central Higgs production.  
 
 The effect of this suppression is often expressed in terms of a ``Survival Probability", $\langle S \rangle$.   In a momentum representation, the  cross section for Higgs production per unit of
rapidity in the central region is
$
\frac{d \sigma_H(s,y_H)}{dy_H} = \frac{ 1\;  }{ \pi^3 (16\pi)^2  s^2 } \int d^2q_{1\perp}  d^2q_{2\perp} |A_H(s,y_H,
q_{1\perp},q_{2\perp})|^2  
$
where $y_H$ is the rapidity of the produced Higgs, $q_{1\perp} $ and $q_{2\perp} $ are transverse momenta of two outgoing fast leading particle in the frame where the momenta of  incoming particles are longitudinal.   ``Survival Probability" is conventionally defined by the ratio
\be
\langle S \rangle\equiv \frac{  \int d^2q_{1\perp}  d^2q_{2\perp} |A_H(s,y_H,
q_{1\perp},q_{2\perp})|^2  }{ \int d^2q_{1\perp}  d^2q_{2\perp} |A^{(0)}_H(s,y_H,
q_{1\perp},q_{2\perp})|^2}
\label{eq:survival}
\ee
where $A^{(0)}_H$ is the corresponding amplitude before eikonal suppression, e.g., given by Eq. (\ref{eq:adsDoublePomeronHiggs}).  
For simplicity, we shall also focus on the  mid-rapidity production, i.e., $y_H\simeq 0$ in the overall CM frame. In this case, $\langle S \rangle$ is a function of overall CM energy squared, $s$, or the equivalent total rapidity, $Y\simeq \log s$.   Evaluating the survival probability as given by (\ref{eq:survival}), though straight forward, is  often tedious. The structure for both the numerator and the denominator is the same. For numerator factor, one has
\bea
&&  \int
 dx_\perp dz \ d\bar z\ P_{13}(z) P_{13}(\bar z) \int d x'_\perp 
 dz' \ d\bar z'\ P_{24}(z) P_{24}(z') \int  e^{i\left(\chi (s,x_\perp-x'_\perp,z,z') - \chi ^*(s,x_\perp-x'_\perp,\bar z,\bar z') \right)}\nn
&&\chi_H(s,s_1, s_2,x^\perp -
x^\perp_H,x'^\perp -x^\perp_H, z,z') \chi_H^*(s,s_1, s_2,x^\perp -
x^\perp_H,x'^\perp -x^\perp_H,\bar z,\bar z')
\eea
where we have made use of that fact that  $z_H\simeq 1/m_H$. To obtain the denominator, one simply removes the phase factor, 
$
e^{i\left(\chi (s,x_\perp,x'_\perp,z,z') - \chi^*(s,x_\perp,x'_\perp,\bar z,\bar z') \right)}
$. It is now clear that it is this extra factor which controls the strength of suppression.

To gain a qualitative estimate, let us consider   the  local limit where $z\simeq \bar z\simeq z_0$ and  $z'\simeq \bar z'\simeq z'_0$, with $z_0\simeq z_0'\simeq 1/\Lambda_{QCD}$. In this limit,  one finds that this suppression factor reduces to
\be
e^{- 2\; {\rm Im}\; \chi (s,x_\perp,x'_\perp,z_0,z'_0)}
\ee
where $ {\rm Im}\; \chi>0$ by unitarity. If follows that, in a super-gravity limit of strong coupling where the eikonal is strictly real, there will be no suppression and the survival probability is 1. Conversely,  the fact that phenomenologically a small survival probability is required is another evidence of we need to work in an intermediate region where $1< j_0<2$.  In this more realistic limit, ${\rm Im}\; \chi$ is large and cannot be neglected.   In particular,  it follows that the dominant region for diffractive Higgs production in pp scattering comes from  the region where
 \be
{\rm Im} \;\;\chi(s,x_\perp- x'_\perp,z,z') = O(1),
\ee
with $z\simeq z' =O(1/\Lambda_{qcd})$. Note that this  is precisely the edge of the ``disk region"  for p-p scattering.   In order to carry out a quantitative analysis, it is imperative that we learn the property of $\chi(s,\vec b, z)$ for $|\vec b|$ large. From our experience with pp scattering, DIS at HERA, etc., we know that confinement will play a crucial role. In pp scattering, since $z\simeq z' =O(1/\Lambda_{qcd})$, we expect this condition is reached at relatively low energy, as is the case for total cross section. It therefore plays a dominant role in determining the magnitude of diffractive Higgs production at LHC.  We will not discuss this issue here further; more pertinent discussions on how to determine $\chi(s,x_\perp- x'_\perp,z,z')$ when confinement is important can be found in Ref. \cite{Brower:2010wf}.

\noindent {\underline{Acknowledgments:}}
We are pleased to acknowledge useful conversations
with M. Block, M. S. Costa,  J. Ellis, E. Gotsman, H. Kowalski, E.  Levin, U. Maor,  D. A. Ross, M. Strassler,  and C. Vergu.  The
work of R. C. B.  was supported by the Department of Energy under
contract~DE-FG02-91ER40676, and that of  C.-IT.  by the
Department of Energy under contract~DE-FG02-91ER40688, Task-A. R.B. and C.-IT. would like to thank the Aspen Center
for Physics for its hospitality during the early phase  of this work. Centro de F\'isica do Porto is partially funded by FCT and the work of M.D. is partially supported by grants PTDC/FIS/099293/2008 and CERN/FP/116358/2010 and by the FCT/Marie Curie Welcome II project.


\begin{thebibliography}{99}
\bibitem{Brower:2006ea}
  R.~C.~Brower, J.~Polchinski, M.~J.~Strassler and C.~I.~Tan,
``The Pomeron and Gauge/String Duality,''
  JHEP {\bf 0712}, 005 (2007).
  
\bibitem{Kharzeev:2000jwa}
  D.~Kharzeev and E.~Levin,
  Phys.\ Rev.\  D {\bf 63}, 073004 (2001)
  [arXiv:hep-ph/0005311].
  
 
  
\bibitem{Brower:2007qh}
  R.~C.~Brower, M.~J.~Strassler and C.~I.~Tan,
``On the Eikonal Approximation in AdS Space,''
  JHEP {\bf 0903}, 050 (2009).
   \bibitem{Brower:2007xg}
   R.~C.~Brower, M.~J.~Strassler and C.~I.~Tan,
 ``On The Pomeron at Large 't Hooft Coupling,''
  JHEP {\bf 0903}, 092 (2009).
  
 
  
  
\bibitem{Cornalba:2006xm}
  L.~Cornalba, et al.,
``Eikonal approximation in AdS/CFT: Conformal partial waves and finite N
four-point functions,''
  Nucl.\ Phys.\  B {\bf 767}, 327 (2007);
``Eikonal approximation in AdS/CFT: From shock waves to four-point
 functions,''
  JHEP {\bf 0708}, 019 (2007);
``Eikonal Approximation in AdS/CFT: Resumming the Gravitational Loop
Expansion,''
{\bf 0709}, 037 (2007);
 ``Eikonal Methods in AdS/CFT: BFKL Pomeron at Weak Coupling,''
 {\bf 0806}, 048 (2008).
 ``Saturation in Deep Inelastic Scattering from AdS/CFT,''
  Phys.\ Rev.\  D {\bf 78}, 096010 (2008).

  
\bibitem{Polchinski:2002jw}
  J.~Polchinski and M.~J.~Strassler,
  JHEP {\bf 0305}, 012 (2003)
  [arXiv:hep-th/0209211].

   
\bibitem{Brower:2010wf}
  R.~C.~Brower, M.~Djuric, I.~Sarcevic and C.~I.~Tan,
  JHEP {\bf 1011}, 051 (2010)
  [arXiv:1007.2259 [hep-ph]];
%
  M.~S.~Costa and M.~Djuric,
  arXiv:1201.1307 [hep-th].

 
\bibitem{Herzog:2008mu}
  C.~P.~Herzog, S.~Paik, M.~J.~Strassler and E.~G.~Thompson,
  JHEP {\bf 0808}, 010 (2008)
  [arXiv:0806.0181 [hep-th]].


\bibitem{Brower:2000rp}
 R.~C.~Brower, S.~D.~Mathur and C.~I.~Tan,
``Glueball Spectrum for QCD from AdS Supergravity Duality,''
  Nucl.\ Phys.\  B {\bf 587}, 249 (2000).
  
 
\bibitem{Brower:2012mk}
  R.~C.~Brower, M.~Djuric and C.~I.~Tan,
  ``Diffractive Higgs Production by AdS Pomeron Fusion,''
  arXiv:1202.4953 [hep-ph].

\end{thebibliography}


\end{document}